\documentstyle[preprint,prl,aps,epsf,12pt]{revtex}

\begin{document}

\renewcommand{\thefootnote}{\fnsymbol{footnote}}

\title{\bf Unusual thermoelectric behavior \\of  packed crystalline granular
metals }

\author{M. Ausloos}

\address{ SUPRATECS, Institute of Physics, B5, University of Li$\grave e$ge,
B-4000 Li$\grave e$ge, Belgium, Euroland}

\author{M. P\c{e}kala}

\address{  Department of Chemistry, University of Warsaw,  Al. Zwirki i Wigury
101, \\PL-02-089 Warsaw, Poland}

\author{J. Latuch}

\address{ Department of Materials Science and Engineering, Warsaw University of
Technology, \\ ul. Wo{\l}oska 141, PL-02-507 Warsaw,  Poland  }

\author{J. Mucha}

\address{ Institute of Low Temperature and Structure Research, Polish Academy of
Sciences, \\ P.O. Box 1410, 50-950 Wroclaw, Poland }

\author{Ph. Vanderbemden}

\address{   SUPRATECS, Institut d'Electricit\'{e} Montefiore B28, University of
Li\`{e}ge, B-4000 Li\`{e}ge, Belgium}

\author{B. Vertruyen and R. Cloots} \address{ SUPRATECS, Department of Chemistry,
B6 University of Li\`{e}ge, B-4000 Li\`{e}ge, Belgium }

\date{\today} \maketitle

\newpage
\begin{abstract} Loosely packed granular materials are intensively studied
nowadays. Electrical and thermal transport properties should reflect the granular
structure as well as intrinsic properties. We have compacted crystalline $CaAl$
based metallic grains and studied the electrical resistivity and the
thermoelectric power as a function of temperature ($T$) from 15 to 300K. Both
properties show three regimes as a function of temperature. It should be pointed
out : (i) The electrical resistivity continuously decreases between 15 and 235 K
(ii) with various dependences, e.g. $\simeq$ $T^{-3/4}$ at low $T$, while (iii)
the thermoelectric power (TEP) is positive, (iv) shows a bump near 60K, and (v)
presents a rather unusual square root of temperature dependence at low
temperature. It is argued that these three regimes indicate a competition between
geometric and thermal processes, - for which a theory seems to be missing in the
case of TEP. The microchemical analysis results are also reported indicating a
complex microstructure inherent to the phase diagram peritectic intricacies of
this binary alloy.

\end{abstract}


\section{Introduction}

\par There are different types of $dense$ granular materials, i.e. those in which
grains are dispersed in a solid matrix in contrast to loosely packed granular
materials. The former class is necessarily a random mixture of two or more phases
with different $I-V$ characteristics. It has been much investigated from a
fundamental point of view along the lines of percolation ideas, for example to
search in the case of superconducting embedded networks \cite{Deutscher} for the
so called critical concentration, whence percolation critical exponents, etc.
\cite{kirkpatrick,clerc}. In the second class, the mixture can be a powder of
even a single chemical type of grains. This class, containing wet or dry
mixtures, is of interest for the technological questions they raise like for
powder, sand or pill packings \cite{Herrmann,Duran}. In between, stem the loosely
packed colloid deposits and polymer like composites \cite{Gadomski1,Gadomski2}.

Depending on the volume fraction they occupy, powders can be more or less dense.
They can be distinguished from a mechanical or optical point of view, i.e.
according to the surface deformation upon loading \cite{Hertz} or to the
modification of the absorbing spectrum \cite{Hunt73,Stroud78,Gerardy1,Gerardy2}
(recall ''invisible'' U2 planes). In the case of electrically insulating grains,
recall that $structural$ properties, e.g. of sand or rice piles are much studied
\cite{Herrmann,Duran}. The $thermal$ properties, like the heat capacity and the
thermal conductivity, e.g. of sand, are of interest as well.

The $electrical$ properties of such systems are also of interest when the grains
are metallic-like \cite{Runge}. It is thought that there is a strong modification
of the interface, when an electric current is imposed, - leading to a strong
modification of the oxide layers formed at the grain surfaces
\cite{Dorbolo1,Dorbolo2,Dorbolo3}, - a modification going even up to welding.
This the Calzecchi-Onesti transition \cite{onesti} leading to sharp transitions,
in the $i-V$ characteristics, when the current reaches a certain value at which
the resistance falls several orders of magnitude.  Microsolderings between grains
may occur before a Calzecchi-Onesti transition \cite{aging}, just like nucleation
sites are initiated by high concentration gradients at a first order phase
transition.  Those solderings are non equilibrium, or irreversible processes
thereby being the fundamental cause for the hysteretic behavior found in $i-V$
curves of granular packing. In the case of not too good conductors, this leads to
the study of hot spots, equally detrimental in superconducting devices
\cite{hotspots}. There are very few papers dealing with the {\it temperature
dependence} of such electrically conductive (loosely) packed materials, - none to
our knowledge about the thermoelectric power temperature dependence.

In huge contrast,  studies of the electrical resistance $R(T)$ of inhomogeneous
composites have been of continuing interest \cite{Abeles,Perenboom}, with much
emphasis on low temperature phenomena, because of superconductivity
\cite{GaboVoit,Angilella,Zeimetz} and tunneling effects; e.g. see
\cite{Choy,Danilov,Efetov,Beloborodov} and references therein. Interestingly, the
thermoelectric power $Q(T)$ has been measured in ceramics, like alumina or SiGe
alloys \cite{alumina,SiGe} and composite superconductors.

It is fair to mention for completeness a few papers on glassy systems
\cite{Qglass1,Qglass2,Qglass3,16,17}, - although these are not truly packed
systems, but are nevertheless close to ... closed packed configurations.

Except for insulating glasses, most composite systems are often made of
electrically conducting grains embedded in an insulator matrix, thus  behave like
a disordered system in the (usually) localized regime. Transport properties
depend whether or not the system is above a percolation transition or not ; they
are usually dominated by carrier hopping $between$ grains, i.e. an $intergrain$
process. Thus they display certain features of the variable-range-hopping (VRH)
phenomenon type observed in {\it doped semiconductors} in fact. In particular,
the electrical resistivity  of such systems has been observed to obey a so called
a stretched exponential dependence, \begin{equation} \rho(T)~=~\rho_0~e^{[
(\frac{T_0}{T})^p]}, \end{equation} where $T$ is the temperature, $T_0$  a
characteristic temperature, and $p~=~1/(d+1)$, in terms of $d$ the dimensionality
of the system \cite{Mott1,Mott2,SE1}.

Yet, it seems obvious that from both structural and chemical points of view, the
$dense$ but loosely packed granular systems should be quite different from doped
materials. In a (loosely or not) packed granular system, the metal (or conducting
entity) occupies a finite volume in space, and hopping from a grain to a (nearest
neighbor) grain is likely to occur

through  rather $wide$ and often $inhomogeneous$ insulating barriers with finite
thickness distributed over a sort of percolating network
\cite{kirkpatrick,Dorbolo1,Dorbolo2,VDBBR,Piasecki}.

In contrast, in the case of a doped material, hopping between two impurity sites
goes through only a $narrow$ size insulating $homogeneous$ barrier.

Therefore, some debate may arise on whether the VRH model of doped materials can
be applied to granular (loosely or densely) compacted metals. The more so for the
thermoelectric power (TEP) or Seebeck coefficient which is a much more elusive
quantity \cite{ziman,DurAuslTEP,Roemer} and does not seem to be found  when
browsing through the VRH literature. Whence the main motivation behind our
studies.

Such electrical and thermal transport properties may be depending on many
conditions, like various types of grain size distributions. This opens up a huge
set of problems, - which cannot be all tackled here. The effective conductivity
and thermoelectric power in such composite systems can be enhanced or reduced, -
the more so as a function of temperature when there are competitive processes. At
this stage it should be pointed out references based on macroscopic points of
view, with considerations based on {\it effective properties} depending on the
type of packing as in
\cite{Stroud75,Bruggeman,Bernasconi,Hori,Nakamura,Bergman,Sangani,McPhedran,Sen,Snarskii}
to quote a few about the (effective medium approximation) electrical
conductivity, while in the case of the thermoelectric power, see
\cite{Skal,Xia,BergmanStroud,BergmanFel}.

In conclusion of the above, it is clear that some attention should be focussed on
the possible competition between geometric and thermal influence with some
subsequent interpretation based on electronic fundamentals.

A simple binary alloy $Ca_{x}Al_{y}$ was selected as an interestingly
representative system. The phase diagram is partially known since 1928 according
to \cite{binaryalloybook}. It was studied when attempting to make crystals from
systems which were initially glassy. Previous investigations did not report any
crystalline phase characteristics, nor {\it a fortiori} temperature dependent
properties of crystalline packed grains.

As described in Sect. 2, an original route has been used for the synthesis of
$CaAl_{2}$; we have obtained tiny crystals  $(0.5 mm)^3$. After compacting the
granular material, we have measured the electrical resistivity ($\rho$),
thermoelectric power ($Q$) and magnetic properties as a function of temperature
(Sect. 3), from 40 to 300 K or so. The temperature dependences of $\rho$ and $Q$
are found to be unusual. The magnetic susceptibility behavior is paramagnetic
(Sect. 3). The finding interpretation relies on temperature dependent percolation
network features, and stresses intergrain and intragrain differences occurring in
such granular packed materials. Band structure considerations from Huang and
Corbett \cite{HuangCorbett} on related $Ca_{x}Al_{y}$, with various $x$ and $y$,
are taken into account in view of interpreting our findings (Sect. 4). The last
section (Sect. 4) serves as much as a place for conclusion as for raising
questions and suggesting further investigations. The chemical characteristics of
the grains are found in Appendix.

\section{Sample Synthesis and chemical characteristics}

\par

Several $Ca_{x}Al_{y}$ phases and stoichiometric compounds are known
\cite{binaryalloybook}. Recently Huang and Corbett \cite{HuangCorbett}
synthesized several of these and characterized them, e.g. $Ca_{13}Al_{14}$ and
$Ca_{8}Al_{13}$, from an electrical resistivity point of view, including some
calculation and discussion of the band structure of both compounds. Both
materials were found to be good electrical conductors, $\rho_{RT}$ $\simeq$ 55
and $\simeq$ 60 $\mu$$\Omega$.cm respectively (with paramagnetic Pauli
properties). Huang and Corbett \cite{HuangCorbett} conclude on a very similar
density of states near the Fermi level.

Our choice of the $x$ and $y$ values for our study has been influenced by
observing the similarity  of the $Ca_{x}Al_{y}$ phase diagram with that of
BiSrCaCuO or  YBaCuO superconducting ceramics, i.e. a peritectic point near $x 
~=~$1 and $y~=~$2, a compound not studied in the crystalline phase to our
knowledge. It is expected that the phase diagram analogy could be of interest for
interpreting features, taking into account the grain intrinsic property, if
necessary.

Master alloys were prepared in an arc furnace under spectrally pure argon
atmosphere with a  0.2 MPa overpressure. Titanium getter was used in order to
eliminate traces of oxygen originating from calcium. The purity of starting
metals was 5N and 3N for Al and Ca, respectively. Samples were homogenized for 15
minutes at 1420 K.

\par The sample is made of very tiny black but shining grains. Even though no
size dispersion distribution was looked for, the size of the grains seems to be
rather uniform, from visual observation, and of the order of $(0.5 mm)^3$. The
samples are   very brittle even after strong packing in a classical press. After
metallographic treatment of the sample surfaces, a grain content is seen through
a high resolution polarized light microscope  (Fig. 1) to be made of dendrites
embedded in a matrix, the dendrite long size roughly ranging between 10 and 50
$\mu$m. Alas it has been impossible to extract the needles from the matrix. The
dendrites are $Al$ rich and made of $CaAl_{2}$ while the matrix is $Al$ poor and
has a composition close to $Ca_{0.6}Al_{0.4}$. The chemical characteristics and a
comment on the grow process, whence the grain structural (unavoidable)
inhomogeneity, can be found in the Appendix. It is shown that the grains are
crystalline.

\section{Measurements}

\subsection{Electrical Resistance}

\par The four-probe method was used for precise measurements of the electrical
resistance of the samples. A dc current up to 10 mA was injected as one second
pulses with successively reversed direction in order to avoid Joule and Peltier
effects \cite{mark12,mark13} and ageing \cite{Dorbolo1}. The electrical
resistivity was calculated from the voltage drop and the sample dimensions, i.e.
7 x 7x 10 mm$^{3}$. The temperature was varied between 20 and 300 K in a closed
cycle refrigerator with stabilization by a LS340 controller with accuracy of 0.01
K. See other details elsewhere \cite{mark14}.

\par The electrical resistance versus temperature variation $R(T)$ measured in a
broad temperature range is shown in Fig. 2 and present three regimes. Different
unusual features  can be observed.  First of all $R$ $decreases$ with e $T$ from
20 K down to 235 K, in particular there is a quasi $linear$ decay between 70 and
235 K. On a log-log plot (Fig. 3) we observe that $R$ remarkably well behaves
like $ \sim T^{-3/4}$ for 15 K$~< T <~$ 70K. $R(T)$  seems to remain rather
constant for higher $T$, or slightly increasing above 235 K. The electrical
resistivity is about $\rho  ~=~ 100 ~\mu\Omega cm$ at room temperature.

\subsection{Thermoelectric power}

\par The thermoelectric power (TEP) has been measured following the method
described in Ref.\cite{mark13}. The imposed temperature gradient is about 1K/cm.
The results are shown in Fig. 4 as $Q$ $vs.$ $T$. $Q(T)$ is positive in the
investigated temperature range and about 20 $\mu V/K$ at 77K and 40 $\mu V/K$ at
room temperature respectively, - values similar to those of poorly metallic
systems. The data presents a bump at ca. 60 K followed by a minimum at 100 K and
a rapid rise thereafter ($\sim ~$ 0.12 $\mu V/K^2$). In Fig. 5, a ln($Q/T$) vs
ln($T$) plot is shown in which a $triple$ regime is well observed. Below 60 K,
$Q(T)/T$ $\simeq ~$ $ 20/\sqrt T$, thus $Q(T)$$\simeq$ 20.0 $T^{0.5}$. In the
intermediary regime, the decay is very regular till 100K. Above 100 K, the slope
is equal to ~-~0.25 on such a log-log plot, indicating that $Q(T)$ behaves like
$T^{3/4}$, thus not quite linearly as it should be usual for metallic systems at
high  $T$.

\subsection{Magnetic susceptibility}

DC magnetic measurements at several temperatures (20 to 300 K) and magnetic
fields (up to 1T) were also performed in a Quantum Design Physical Property
Measurement System, when it worked,  using an extraction method.

\par  The response of the material is that of a paramagnetic one over the whole
temperature range investigated, with a susceptibility of the order of 8.6
$10^{-6}$ (SI units) at $T~ = ~$300K. It is quasi independent of temperature :
the susceptibility decreases by a factor of 3 between 50 K and 300K, in a field
of 1 T, but  this is not displayed here.

\section{Discussion}

The discussion is essentially based on the microstructure optically and
microscopically observed, i.e. a packed mixture of (apparently $heterogeneous$,
see Appendix) tiny grains, together with the unusual temperature dependence of
the electrothermal property for stressing the features and emphasizing
fundamental considerations.

First, let it be observed that even though each grain is surely multiphasic, see
Appendix, both phases, according to Huang and Corbett \cite{HuangCorbett} have
equivalent electrical (metallic) properties. Therefore for discussing the
electrical resistivity, at least, we can safely assume that each grain is
monophasic and metallic. If the grains were made of insulating components in a
not well conducting matrix, thus having a peculiar heterogeneity as found in
metal-insulator composites, one might argue that charging effects on inner
surfaces might complicate the analysis. This does not seem to be the case here,
whence giving some validity for effective medium theory considerations in
describing  each grain.

One might next wonder whether the sample network structure could be much modified
as a function of temperature, leading to  a temperature dependent
(macrostructural) transition. Even though we cannot entirely disregard such
effects due to grain thermal expansions, we assume that the granular network
structure, whence the number of contacts should be rather stable. However the
intergrain connections could be temperature dependent. Therefore the electrical
conduction process should be  only resulting from a intergrain hopping  process
and an intragrain (metallic like) conductivity. Both having a  different
temperature dependence. Again, an effective medium theory should be valid.

\subsection{Electrical Resistance}

Various mechanisms are invoked when explaining a decreasing  $R(T)$ behavior: (i)
an Arrhenius excitation, i.e. an $exponential$ decrease corresponding to a
thermal barrier leakage when the temperature increases; (ii) a Mott, so called
variable range hopping (VRH) \cite{Mott1,Mott2} mechanism, or (iii) a
Shklovskii-Efros (SE) effect \cite{SE1,SE2}, i.e. a small Coulomb gap opens up at
the Fermi surface as a result of Coulomb interactions between the quasiparticle
carriers, thereby modifying Mott VRH result. In these cases the electrical
resistivity can be expressed as a stretched exponential \begin{equation}
\rho(T)~=~\rho_0 \quad e^{[ (\frac{T_0}{T})^p]}, \end{equation} where $\rho_p$ is
a material weakly $T$ dependent quantity, and $p~=~$1/4 or 1/2 for the Mott or SE
type conductivity; $T_0$ characterizes the variable range hopping energy scale.
The classical Arrhenius law corresponds to $p$=1. Plots of log $\rho$ vs.
$T^{-1/2}$  or $T^{-1/4}$ look so far from any straigtht line that they are not
displayed; they do $not$ indicate any (relatively extended) linear region which
could convince us of arguing in favor of a Mott or SE mechanism in some range. In
fact it is found from Fig. 6, that below 70 K \begin{equation} \rho(T)~=~
\hat{\rho_0} \quad(\frac{T }{77})^{-3/4}, \end{equation} with $\hat{\rho_0} ~ =~
25 ~\mu\Omega cm$ when $T$ is measured in Kelvin.

Unusual fractional exponents are often interpreted in other condensed matter
research through fractal considerations for the underlying (geometrical) network
supporting the elementary excitations of interest \cite{Orbach}, - themselves
having a possible temperature dependence (which scales like their intrinsic
interaction energy). Recall here as an analogy the case of magnetic disordered
sytems, characterized by some exchange energy $J$ (and some directly related)
critical temperature $T_c$, on a (random) network, characterized by some fractal
dimension. The best description of those systems is in fact \cite{staufferbook}
through two different coherence lengths, $\xi_g$ and $\xi_T$, one (usually
temperature independent) for the network, one  (strongly temperature dependent)
for the magnetic structure.

The present electrical conduction process with the temperature behavior of
$\rho(T)$, i.e. the $T^{-3/4}$ law is best interpreted in terms of a similar
argument covering the whole temperature range. In particular, the power law
($\sim T^{-0.75}$) resistance decrease can be thought to result from a $thermal$
effect, on a resistive (geometrically disordered) backbone, not changing with
temperature in this low $T$ range. Indeed in presence of such a geometric
disorder, the charge carrier mean free path $\ell$ can be considered in the usual
linear superposition approximation as \cite{ziman} \begin{equation} \label{ell}
\frac{1}{\ell}~=~\frac{1}{\ell_T}+\frac{1}{\ell_{g}} \end{equation} where
$\ell_T$ and $\ell_{g}$ are the mean free path of carriers respectively due to
(grain size limited) thermal effects and  intrinsic geometric disorder. Moreover,
in this less conducting, i.e. low temperature regime, Orbach description of
random media excitations, \cite{Orbach} can be used to interpret the exponent
$3/4$ as the signature of the percolation network for the hopping charge
carriers, - whence finding  a (very reasonable) effective (fractal-like
\cite{Orbach,staufferbook},) dimensionality $\tilde{\delta}  ~=  ~9/4$ of the
{\it three dimensional}, since $\tilde{\delta}>2$, backbone \cite{tunnelgrain}.

The 60-70 K break (or crossover) indicates the energy range at which the thermal
process takes over on the geometric disorder. From a microscopic point of view,
this temperature value allows us to estimate the (average) electrical carrier
energy distance between the Fermi level and the bottom (observe that $Q(T) >0$,
in Fig. 7) of the conduction band. This hopping (or barrier) energy being
overcome  at such a temperature, leads to a more easy conduction process, whence
a weaker $\rho$. Moreover, above such a $T$, in the so called $intermediate$
temperature range the $R(T)$ linear decay reminds us of the behavior found in
heavy fermion systems. Indeed an electrical resistivity decreasing with
temperature, as in Fig. 5, is also observed in heavy fermion materials
\cite{ladek,Coqblin,Ginzburg}. Such a feature is associated to a sharp DOS near
the Fermi level and carrier (de)localization \cite{Lee}.

Finally, at higher temperature,  the smooth decay of $R(T)$ with increasing $T$
can be attributed to the Fermi-Dirac function behavior increasing width as $T$
increases and further charge carrier delocalization \cite{Lee}. The electron (or
rather hole, here) band mass contains a more important imaginary contribution as
a function of temperature, and the charge carrier wave function spreads out. The
$intergrain$ thermally activated charge (ballistic) hopping process. becomes more
pronounced  at high temperature and competes with ($p$-like) metallic conduction
in the grains. Recall a similar type of mechanism due to (dense) granular
structure in mixed superconductors \cite{Deutscher,Efetov,Khan} like Al-Ge
mixture, at ''not very low temperatures'' \cite{Gerber} where the $R(T)$ decay is
like $T^{-~0.117}$, an exponent  similar but quite different from $3/4$. Assuming
that all (weak) localization effects are suppressed by the temperature the
occurrence of a phase transition in $R(T)$ exists depending on the tunneling
conductance $g$, and the phase coherence on both sides of the tunneling junction.
For such systems, at small $g$ the conductivity $exponentially$ decays with
temperature, but has a $logarithmic$ temperature dependence at large $g$. This
(effective) conductance picture can be used analogously here, since the
intergrain barrier conductance $g$ likely increases with temperature. However the
temperature dependences found here above are quite different from those in the
granular metals discussed by  e.g. Efetov and Tschersich \cite{Efetov}. In fact,
in view of the different temperature dependences,  we prefer to consider that the
effects are due to intrinsic density of states rather than mobility constraints
at not too low temperature.

\subsection{Thermoelectric power}

Turning to the TEP data, we have already argued elsewhere \cite{kdurma} that $Q$
should be considered as a transport coefficient and not assimilated only to the
change in entropy or density of states \cite{ziman}. Nevertheless the above
recalled considerations on mean free path selection for the electrical charge
carrier scattering processes are also very useful in order to interpret the
$Q(T)$ behavior at low $T$. As in an effective medium approximation, the behavior
of $Q(T)$ depends on the dominant scattering process. Notice that the concept of
a percolation network made of tunneling barriers  with a $distribution$ of
temperature gradients is not usual nor easily simulated for TEP. However see some
similar consideration by Danilov et al. in \cite{tunnelgrain}.

Let us also recall that a TEP measurement implies no external electrical current.
Whence it is unlikely that some ''barrier ageing'' or ''hot spots'' take place in
the temperature range of interest through a Calzecchi-Onesti-like transition
\cite{onesti}. Moreover local charging can also be neglected, in view of the
metallic nature of the grains, - though local hot spots might exist. To our
knowledge they have not been observed in such configurations. Thus within the
above line of thought, both increase in charge and heat transport as a function
of temperature and the large thermoelectric effect at room temperature can also
be understood to result from a delocalization process on the complex barrier
network, with two competing characteristic mean free paths, and a wakening of the
contact TEP due to Fermi surface widening with temperature. There is to our
knowledge no such a theory for TEP.

To be fair one might also argue on whether the $N$-shape of $Q(T)$ is not due to
a phonon drag. Our argument to dismiss phonon drag in favor of competing carrier
localization lengths stems in the fact that phonon drag is usually associated to
long wave length phonons and are thus sensitive mainly to boundary scattering. In
view of the grain size and granular packing, it could be the case here. However,
phonon drag description leads to a predicted behavior like $T^3$ for metallic
systems, -  this is not found here at all. IAs an argument, recall that TEP of
compacted small size (1 mm) homogeneous Ge grains has been previously studied at
low temperature, without any convincing evidence that the behavior was due to
phonon drag \cite{Frederikse,Geballe,Herring}. In fact, a $Q\sim T^{-9/2}$ should
be expected \cite{Nolas}, a very different behavior from what is observed here.
Whence it can be discounted that the peak in TEP is due to phonon drag processes.
We thus propose that the features are due to an intrinsically competitive
(geometric-thermal) dissipation mechanism, involving intra and intergrain
parameters. Both a change in  DOS and/or electronic heat conduction scattering
should be taken into account in theoretical work \cite{ziman}.

\section{Conclusion}

After studying  the effect of electrical currents on  intergrain barrier ageing,
at a fixed (room, in fact) temperature  for loosely but dense packing
\cite{Dorbolo1,Dorbolo2,Dorbolo3}, one step, among many others, consists in
studying  the temperature depedence of basic transport properties, like the
electrical resistivity and the thermoelectric power. One might wonder whether
features are due to the packing structure only or  whether the electrothermal
properties are sensitive to grain contacts.

We have synthesized a packing of grains made of well identified crystalline
phases \cite{HuangCorbett}. The microstructure and chemical content of the tiny
grains have been examined. We have measured the electrical resistivity and the
thermoelectric power.

The  thermoelectric power shows three regimes, like the electrical resistivity, -
in both cases within identical temperature intervals. Readily the electrical
resistivity has a long unusual power law decreasing regime between 20 and 200 K.
In particular, the thermoelectric power shows an unusual square root temperature
dependence at low temperature. It seems that the observed complex dependences can
be considered to result from a competition between metallic intragrain and
semiconducting intergrain barrier properties.

From our observations and reported features by others, we can neglect the
inhomogeneity of the grains when discussing electrical and thermal transport
properties. The system is thus thought to be described as a network of resistive
packed ''macro'' grains having boundaries well connected between conducting
grains. The role of the density of carriers in the   grains is emphasized. Indeed
in  view of the rather high value of the overall resistance, even taking into
account the effect of ''macro'' grain barriers, a sharp density of states (DOS)
at the Fermi level, close to some conducting gap, maybe expected, suggesting some
localization. Thus such electrical and thermal transport properties indicate a
competition between mean free paths, later overcome by easy electronic tunneling.

\par We stress that the corresponding features in the temperature regimes both
for $R(T)$ and $Q(T)$ allows for  strong arguments on the physics of such packed
granular metallic material properties. The $R(T)$ behavior, $\rho\sim T^{-3/4}$
at low temperature followed by a linear decrease being unusual for good metallic
systems is interpreted as resulting from a competition process between intragrain
and intergrain conductivity.  At low temperature we suggest an intergrain
limiting conductivity with intragrain electronic localization and intergrain
mediated hopping. An increase in temperature leading to a (low) saturation and
delocalization regime.

From the resistance data break we can even conjecture the existence of a small
gap ($\sim$ 60 K, between the $E_F$ level and the top of the valence band, in
$CaAl_2$, such that thermal excitations $reduce$ the number of truly conducting
electrons by exciting them into a narrow (heavy mass) conduction band. This
suggests the great interest in calculating the band structure of such
compound(s); this could be complemented by specific heat and optical
measurements.

Second it is necessary to insist about the unusual square root dependence of the
thermoelectric power at low temperature, $Q\sim T^{1/2}$, and $Q\sim T^{3/4}$ at
high temperature. Under the conjecture that the behavior is consistent with an
electronic hopping between grains at low temperature, we thus report for the
first time the behavior of a charge carrier on a percolation network under a
temperature gradient. A search of the literature has not revealed whether this
TEP behaves   like that in glasses \cite{Qglass1,Qglass2} or cermets, nor
indicates a behavior controlled by grain size effects \cite{Nolas}. Therefore in
agreement with the above considerations on $R(T)$ it may be concluded that such a
behavior represents (as observed for the first time maybe here) a {\it granular
conducting medium thermoelectric power}.

An intersting point ot be raised here is whether a Peltier effect and a Seebeck
effect are equivalent  \cite{Anatychuk} in granular materials, e.g. would hot
spots and aging occur, leading to some hysteresis, if rather than a temperature
gradient (Seebeck configuration) a potential difference (Peltier configuration)
is applied, and the temperature gradient next measured in order to obtain the TEP
characteristics. This is an open question.

In conclusion, there is often some question about whether single crystals are the
only samples of fundamental interest, to be studied and reported upon. No doubt
that they easily bring some information on microscopic parameters. Yet, it is
also often shown that polycrystalline samples are useful when fine measurements
are performed. Moreover, properties of packed systems receive some attention
nowadays. To explain their properties become a challenge in condensed matter. In
fact we  report an unusual decay of the electrical resistance with temperature, a
square root of temperature dependence for the TEP, at low $T$ and some consistent
temperature ranges where phenomena can be distinguished due to   microscopic
processes in metallic grains.

\vskip 0.5cm  \newpage \vskip 0.5cm

{\bf Appendix}

An EDX analysis of the synthesized grains indicates that the dendrites (Fig. 1)
are $Al$ rich and made of $CaAl_{2}$ while the matrix is $Al$ poor and has a
composition close to $Ca_{0.6}Al_{0.4}$, - in fact a mixture of $Ca_{13}Al_{14}$
and $Ca_8Al_{3}$, taking into account the Huang and Corbett report
\cite{HuangCorbett} and  a very recent \cite{Ozturk} phase diagram report
sketched in Fig. 6. An X-ray diffraction analysis (Fig.7) proves that there are
in fact three crystalline phases. For identifying the $Ca_{8}Al_{13}$ and
$Ca_{13}Al_{14}$  data from JCPDS files were used.  From the X-ray data, the
lattice parameter of the $CaAl_{2}$ phase is found to be $a=0.8.045$ nm, larger
than the one reported as equal to $a=0.8038$ nm  in ref. \cite{binaryalloybook}.
According to JCPDS 75-0875,  $ a= 8.02 \AA$ with Fd-3m as space group. The
presence of three phases  likely indicates some slight $Ca$ excess in the initial
powder.

A comment is here in order for explaining the growth process, - indeed somewhat
analogous to the superconducting ceramics mentioned above. Starting from a
stoichiometric powder composition close to $CaAl_{2}$, we consider that the
solidification (quenching) process induces the appearance of $CaAl_{2}$ grains
embedded in a fluid phase. Under further cooling the latter solidifies,
$Ca_{13}Al_{14}$ precipitates near (or on)  $CaAl_{2}$ grains and decomposes at
the 850 K peritectic temperature into a $Ca_{13}Al_{14}$ and a $Ca_{8}Al_{3}$
composite matrix leading to the microstructure schematically described in Fig. 8.
In so doing one can understand that large $CaAl_{2}$ dendrite grains may be
covered by a tiny $Ca_{13}Al_{14}$ layer, all embedded in a mainly
$Ca_{0.6}Al_{0.4}$ matrix containing tiny, - less than 1 $\mu$m size from optical
observation, $Ca_{8}Al_{3}$ and $Ca_{13}Al_{14}$ precipitates. It is interesting
to notice that the grains are made of needles in a matrix, in contrast to the
rather spherical 211 particles in the 123-YBCO matrix.

\vskip 0.5cm \par \par {\large \bf Acknowledgements}

\indent Work supported in part by the University of Li\`{e}ge Research Council,
by PST.CLG.977377 and by KBN grants 7T08A 02820 and 2PO3B12919.

\begin{figure}[htb]  \caption{ Polarized light microscopy analysis of a grain of
the investigated system, showing $CaAl_{2}$  dendrites, in a $Ca_{14}Al_{14}$ -
$Ca_8Al_{3}$ matrix; the size of the photography diagonal is about 100$\mu$m}
\end{figure}

\begin{figure}[htb]   \caption{ Electrical resistance vs. temperature of a
polycrystalline dense packed sample based on $CaAl_{2}$; the sample dimensions
are 7 x 7x 10 mm$^{3}$, leading to a resistivity at room temperature $\simeq 100
\mu\Omega$cm} \end{figure}

\begin{figure}[htb]  \caption{ Same as on Fig. 2, on a log-log plot, with  fit to
a line with slope=-3/4 at low temperature} \end{figure}

\begin{figure}[htb]  \caption{ Thermoelectric power $Q(T)$ vs. temperature $T$ of
a polycrystalline dense packed sample showing a $N$-shape behavior} \end{figure}

\begin{figure}[htb]  \caption{ Same as on Fig. 4 but for $Q(T)/T$ vs. temperature
$T$  on a log-log plot, with fits at low and high temperature} \end{figure}

\begin{figure}[htb]  \caption{ Phase diagram of the binary system $CaAl$,
extracted from ref. [76] } \end{figure}

\begin{figure}[htb]  \caption{ X-ray diagram of a dense as synthesized $CaAl$
system; the inset points out to peaks belonging to the so called matrix phase}
\end{figure}

\begin{figure}[htb]  \caption{ Sketch of the observed microstructures in relation
to the phase diagram of the binary compound $CaAl$ with emphasis on the
peritectic temperature region} \end{figure}

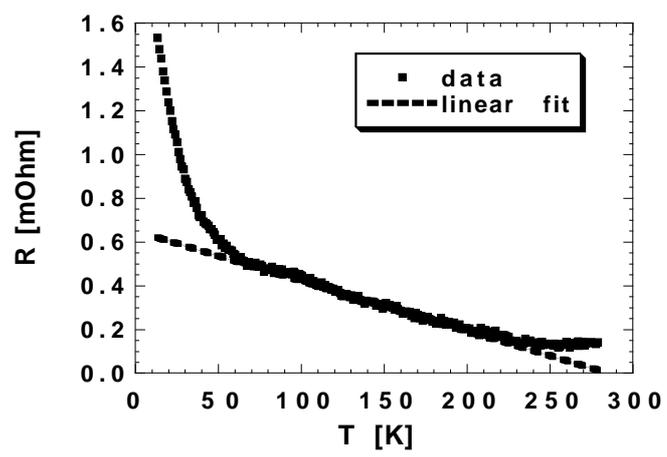
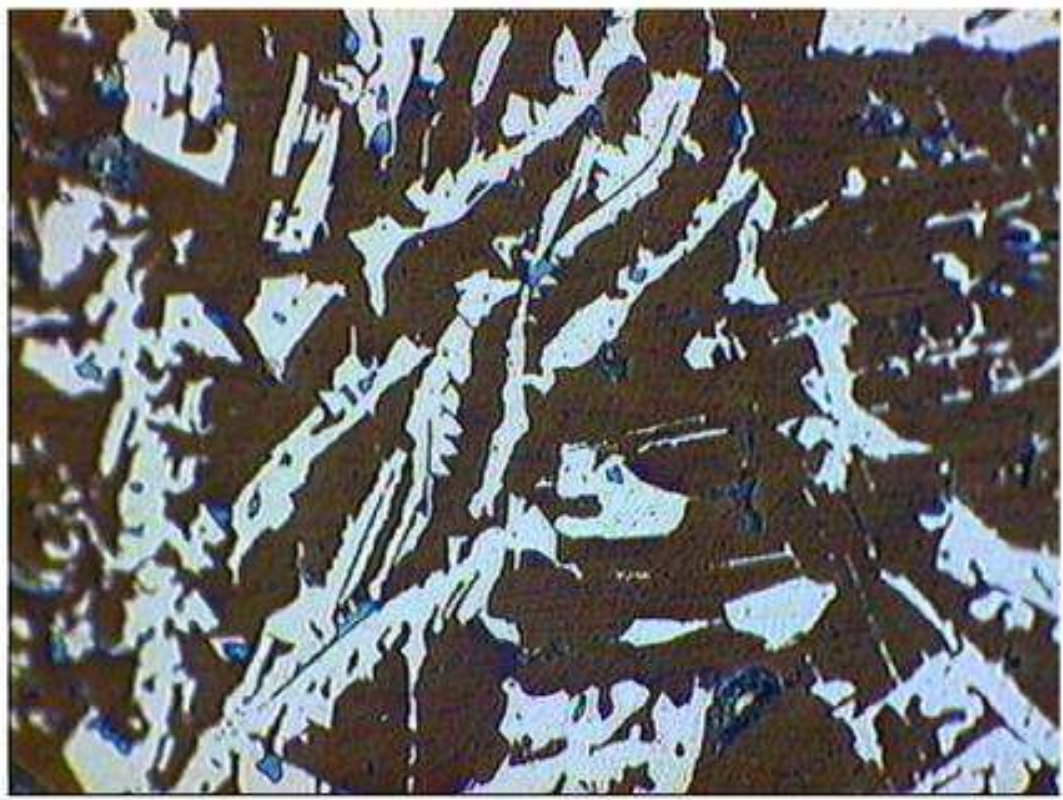
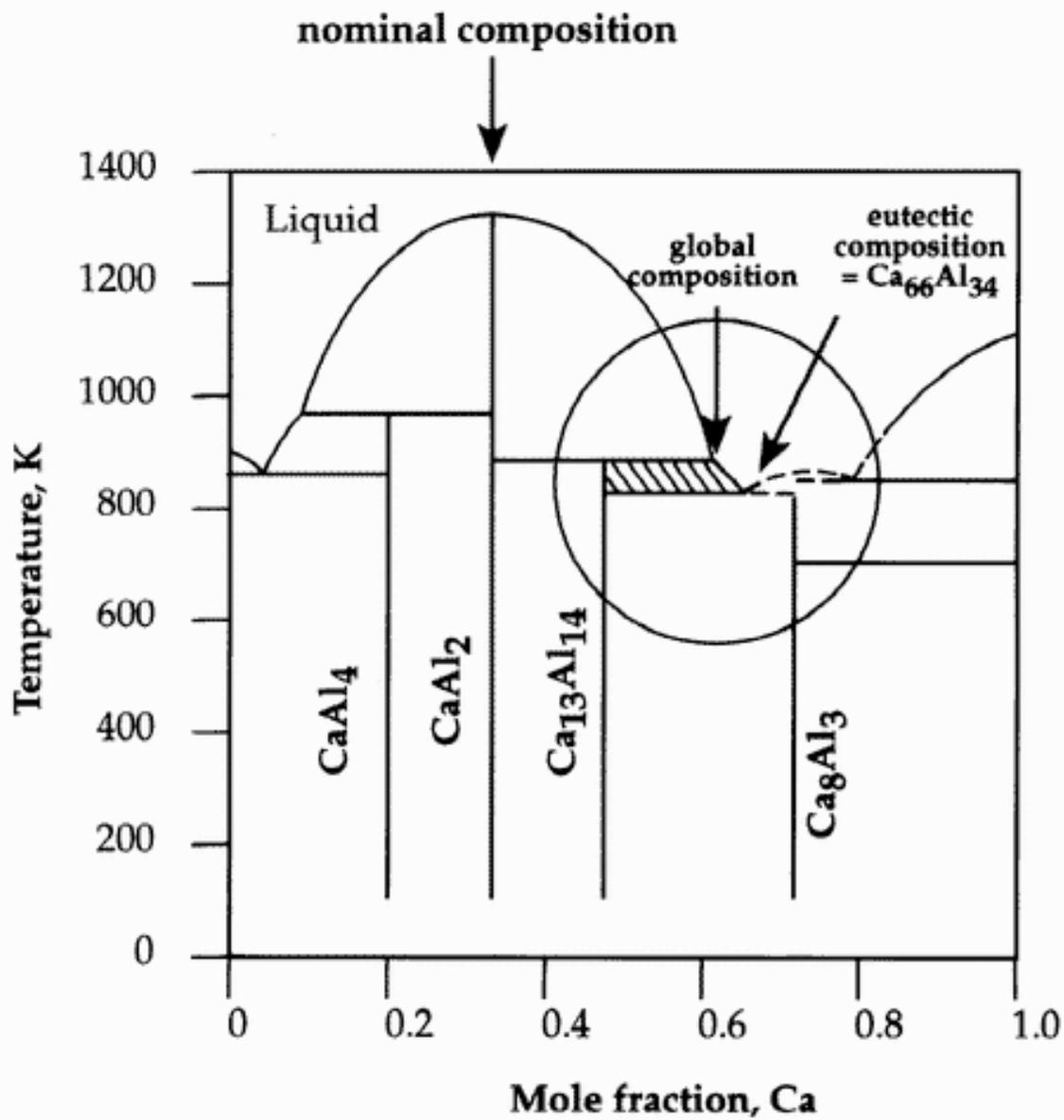
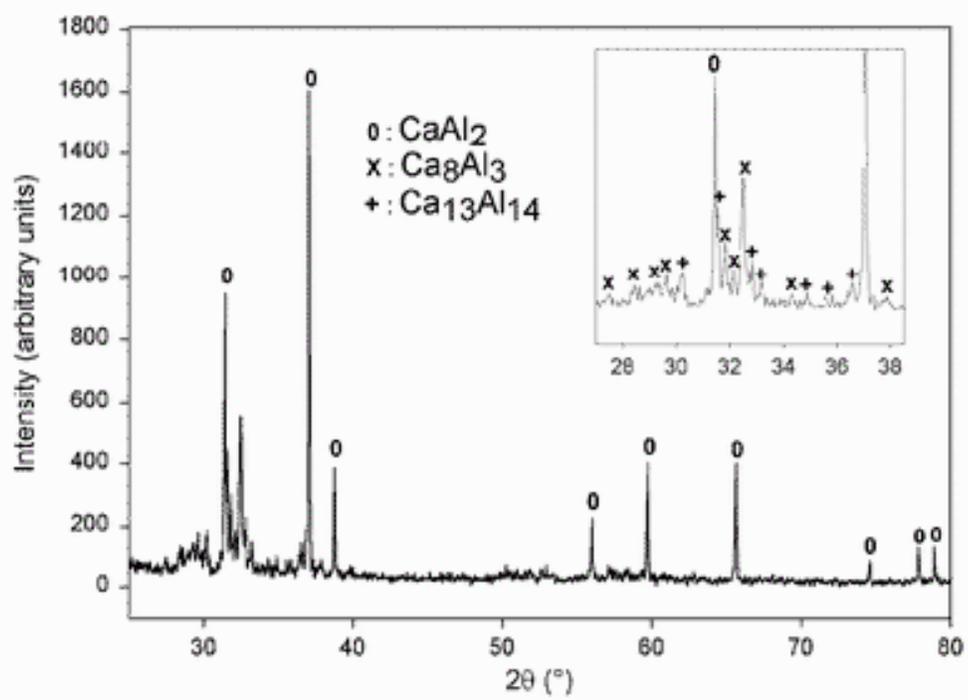
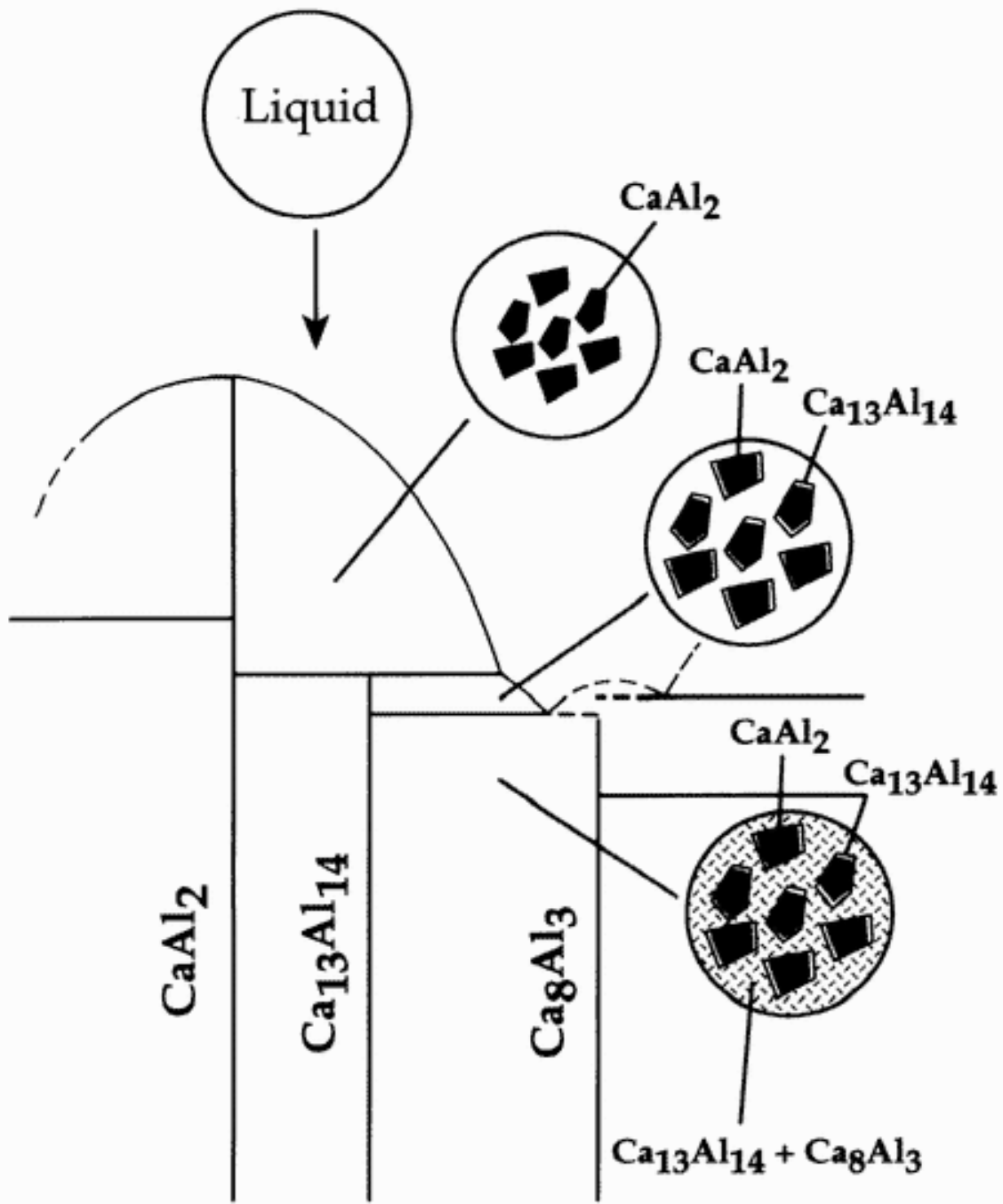
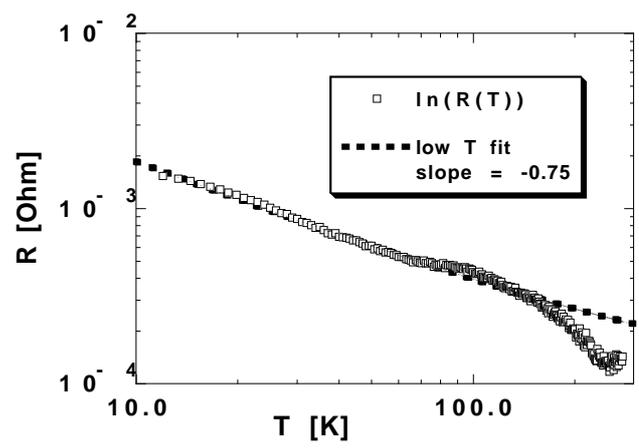
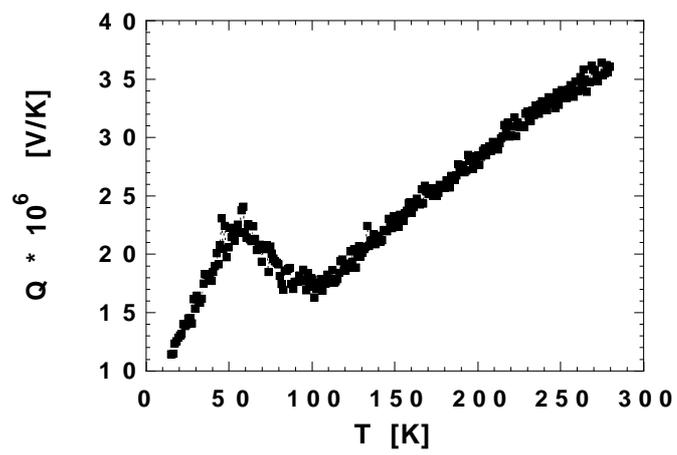
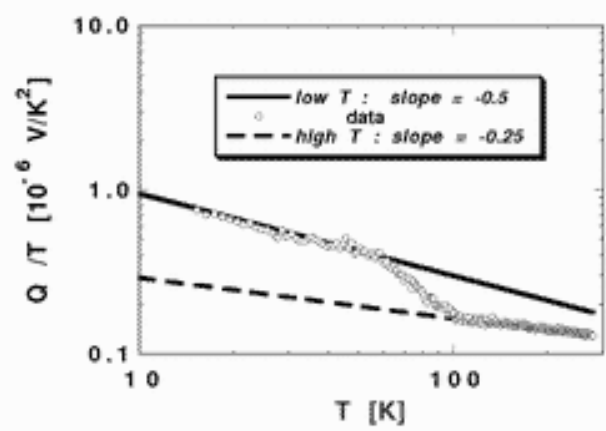
\end{document}